\documentclass[usenatbib]{mn2e}

\usepackage{amsmath}

\title{A variational formalism for tidal excitation: non-rotating, homentropic stars}
\author[Yasser Rathore, Avery E. Broderick and Roger Blandford]{Yasser Rathore,\thanks{yasser@caltech.edu; aeb@caltech.edu; rdb@tapir.caltech.edu} Avery E. Broderick\footnotemark[1] and Roger Blandford\footnotemark[1] \\ Theoretical Astrophysics, Caltech 130-33, Pasadena, CA 91125, USA}
\date{Accepted 2002; Received 2002}
\pagerange{\pageref{firstpage}--\pageref{lastpage}}
\pubyear{2002}

\begin{document}
\maketitle
\label{firstpage}
\begin{abstract}
We present a variational formalism for describing the dynamical evolution of an oscillating star with a point-mass companion in the linear, non-relativistic regime. This includes both the excitation of normal modes and the back-reaction of the modes on the orbit. The general formalism for arbitrary fluid configurations is presented, and then specialized to a homentropic potential flow. Our formalism explicitly identifies and conserves both energy and angular momentum. We also consider corrections to the orbit up to 7/2 post-Newtonian order.
\end{abstract}
\section{Introduction}
When a star orbits a companion, there can be a resonant excitation of stellar oscillations if the orbital period becomes close to a multiple of the oscillation period. This can happen as a result of gravitational radiation or interaction with a third body, for example. This is a deceptively complex dynamical system because there will be a back-reaction of the oscillation on the orbit. In this paper we consider a variational approach to treating this problem.

Variational principles for linear, adiabatic, non-radial stellar oscillations have been considered by several people \citep[e.g.][]{cha64,lyn67}. Subsequently, a variational approach to the excitation of tides in binary systems was considered by \citet{Gin80}. However, their description was limited to polytropic equations of state, and did not account for orbital evolution due to general relativistic effects which are significant for the long-term evolution of systems involving compact stars. In this paper, we consider a more general variational approach to tidal excitation. Although we limit our detailed discussion to irrotational flow, the formalism is presented in a way so as to allow generalisation to more complicated flows.

In \S 2, we provide an overview of one variational approach to the dynamics of perfect fluids. We then derive the most general Lagrangian for tidal excitation, and consider the special case of homentropic, irrotational flow. In \S 3, we derive the equations for tidal excitation in terms of the normal mode amplitudes, and obtain expressions for the conserved energy and angular momentum. We also consider corrections to the orbit due to general relativistic effects. In \S 4, we discuss the applicability of our formalism. Finally, in \S 5, we reprise our conclusions. 
\section{The Lagrangian}
\subsection{Overview of variational fluid mechanics}
In this section, we outline some of the important results from a variational formulation of fluid mechanics. Our discussion follows the review by \citet{sal88} where a more detailed exposition may be found. An important difference is that we have extended the formalism to accommodate a self-gravitating fluid in a non-inertial reference frame.

The simplest variational formulation is to use a continuum version of the Lagrangian from classical mechanics. In this approach, the Lagrangian for the fluid is just the classical Lagrangian for a system of particles distributed continuously in space. Let $\bld x(\bld a,\tau)$ be the position, relative to the centre-of-mass, of the fluid particle identified by the labeling coordinates $\bld a$ at time $\tau$. We shall distinguish between the time coordinates $\tau$ and $t$. These are equal in magnitude, but partial derivatives with respect to $\tau$ are at constant $\bld{a}$, whereas those with respect to $t$ are at constant $\bld{x}$---in other words, $\partial/\partial\tau$ corresponds to a convective derivative. There is considerable freedom in the choice of labeling coordinates, but it is convenient to choose them so that they are related to the mass density of the fluid by
\begin{equation} \label{eq:labcoords}
\rho = \frac{\partial (\bld a)}{\partial (\bld x)}.
\end{equation}
It should be noted that this just corresponds to the choice of a constant mass for the fluid particles (i.e.~the mass density of the fluid is directly proportional to the number density of particles). This has the advantage that mass conservation is implicit in our choice of $\bld a$, as can be verified by a direct application of $\partial/\partial\tau$ to (\ref{eq:labcoords}):
\begin{equation} \label{eq:masscons}
\frac{\partial\rho}{\partial\tau} + \rho\nabla\cdot\bld{u} = 0,
\end{equation}
where $\bld{u}\equiv\partial\bld{x}/\partial\tau$, $\nabla$ is the gradient operator in $\bld{x}$-space, and we have used the fact that the inverse of a matrix $A$ may be written as
\begin{equation} \label{eq:matrixinv}
A_{ji}^{-1} = \frac{\partial\ln ||A||}{\partial A_{ij}},
\end{equation}
where $||A||\equiv\det (A)$.
We can now write down the Lagrangian for the fluid as
\begin{equation} \label{eq:flulag1}
\begin{split}
L_* = \int\! d\bld{a}\ \Bigg[ &\frac{1}{2}\left(\frac{\partial\bld{R}_*}{\partial\tau}+\frac{\partial\bld{x}}{\partial\tau}\right)^2 \\ &- E\left(\frac{\partial(\bld{x})}{\partial(\bld{a})},S(\bld{a})\right) - \Phi(\bld{x}) \Bigg],
\end{split}
\end{equation}
where $\bld{R}_*$ is the location of the centre-of-mass, $\Phi$ is the potential for external forces, and $E$ is the specific internal energy which is a prescribed function of the specific volume $\rho^{-1}$ and the specific entropy $S$. Note that $S$ depends only on the labeling coordinates $\bld{a}$. This is, in essence, the perfect-fluid approximation:
\begin{equation*}
\frac{\partial S}{\partial\tau} = 0.
\end{equation*}
It can be shown that the variation with respect to $\bld{x}$ of (\ref{eq:flulag1}) yields the Euler equation (see Appendix \ref{app:euler}).

For computational purposes, it is more convenient to rewrite (\ref{eq:flulag1}) in Eulerian form. This is straightforward to accomplish by noting that the time-dependent map \mbox{$\bld{x}=\bld{x}(\bld{a},\tau)$} uniquely determines the inverse map $\bld{a}=\bld{a}(\bld{x},t)$. Therefore, the requirement that the action be stationary under arbitrary variations $\delta\bld{x}$ in the forward map is equivalent to the requirement that the action be stationary under variations $\delta\bld{a}$ in the inverse map. After dropping two total time derivatives, we can now write the fluid Lagrangian as
\begin{equation} \label{eq:flulag2}
\begin{split}
L_* &= \frac{1}{2}M_*\dot{\bld{R}}_*^2 \\ &
\begin{split} + \int\! d\bld{x}\ \bigg\{ &\rho\left[ \frac{1}{2}\bld{u}\cdot\bld{u} - E\left(\rho,S(\bld{a})\right) - \Phi(\bld{x}) \right] \\ &+ \frac{\partial\rho}{\partial t}\bld{x}\cdot\dot{\bld{R}}_* \bigg\},
\end{split}
\end{split}
\end{equation}
where $M_*$ is the total mass of the fluid. However, before we can consider the variation with respect to $\bld{a}$ of (\ref{eq:flulag2}), we must express the velocity $\bld{u}$ as a function of $\bld{a}$ and its derivatives. Alternatively, we can include the relevant relations as constraints in the Lagrangian and then vary $\bld{u}$ and $\bld{a}$ independently. The required relations are given by
\begin{equation}
0 = \frac{\partial\bld{a}}{\partial\tau} = \frac{\partial\bld{a}}{\partial t} + \left(\bld{u}\cdot\nabla\right)\bld{a}
\end{equation}
which are the so-called \emph{Lin constraints}. We may also include mass conservation (\ref{eq:masscons}) as a constraint in the Eulerian form (\ref{eq:flulag2}) of the Lagrangian. This gives us
\begin{equation} \label{eq:flulag4}
\begin{split}
L_* &= \frac{1}{2}M_*\dot{\bld{R}}_*^2 \\ &
\begin{split} + \int\! d\bld{x}\ \bigg\{ &\rho\left[ \frac{1}{2}\bld{u}\cdot\bld{u} - E\left(\rho,S(\bld{a})\right) - \Phi(\bld{x}) - \bzeta\cdot\frac{\partial\bld{a}}{\partial\tau} \right] \\ &+ \phi\left[\frac{\partial\rho}{\partial t} + \nabla\cdot\left(\rho\bld{u}\right)\right] + \frac{\partial\rho}{\partial t}\bld{x}\cdot\dot{\bld{R}}_* \bigg\},
\end{split}
\end{split}
\end{equation}
where $\bzeta$ and $\phi$ are Lagrange multipliers, and we now consider the independent variations $\delta\bld{u}$, $\delta\bld{a}$, $\delta\bzeta$, $\delta\rho$, and $\delta\phi$.

The $\bld{u}$ variation of (\ref{eq:flulag4}) yields
\begin{equation} \label{eq:genvel}
\bld{u} = \zeta_i\nabla a_i + \nabla\phi,
\end{equation}
which can be used to eliminate $\bld{u}$ from the Lagrangian. After dropping a time derivative and integrating one term by parts, (\ref{eq:flulag4}) becomes
\begin{equation}
\begin{split}
L_* &= \frac{1}{2}M_*\dot{\bld{R}}_*^2 \\ &
\begin{split} - \int\! d\bld{x}\ \bigg\{ \rho\bigg[ &\bzeta\cdot\frac{\partial\bld{a}}{\partial t} + \frac{\partial\phi}{\partial t} + \frac{1}{2}\bld{u}\cdot\bld{u} \\ &+ E\left(\rho,S(\bld{a})\right) + \Phi \bigg] - \frac{\partial\rho}{\partial t}\bld{x}\cdot\dot{\bld{R}}_* \bigg\},
\end{split}
\end{split}
\end{equation}
where $\bld{u}$ is now just an abbreviation for (\ref{eq:genvel}), and the independent variations $\delta\bld{a}$, $\delta\bzeta$, $\delta\rho$ and $\delta\phi$ are to be considered.

So far we have neglected the effects of self-gravitation, considering $\Phi$ to be an externally imposed potential. We may now incorporate self-gravity in our formalism by including the Poisson equation. Our most general perfect-fluid Lagrangian is then
\begin{equation} \label{eq:flulag5}
\begin{split}
L_* &= \frac{1}{2}M_*\dot{\bld{R}}_*^2 \\ &
\begin{split} - \int\! d\bld{x}\ &\bigg\{ \rho\bigg[ \bzeta\cdot\frac{\partial\bld{a}}{\partial t} + \frac{\partial\phi}{\partial t} + \frac{1}{2}\bld{u}\cdot\bld{u} + E\left(\rho,S(\bld{a})\right) \\ &+ \Psi + \Phi \bigg] + \frac{1}{8\pi G}\nabla\Psi\cdot\nabla\Psi - \frac{\partial\rho}{\partial t}\bld{x}\cdot\dot{\bld{R}}_* \bigg\},
\end{split}
\end{split}
\end{equation}
where $\Psi$ is the self-gravitational potential, and the independent variations are $\delta\bld{a}$, $\delta\bzeta$, $\delta\rho$, $\delta\phi$, and $\delta\Psi$.
\subsection{Coupling to a point-mass}
Consider now the problem in which the external potential $\Phi$ is due to the gravitational field of a point-mass $M_h$. The Lagrangian for the whole system is therefore
\begin{equation} \label{eq:lag1}
\begin{split}
L &= \frac{1}{2}M_h\dot{\bld{R}}_h^2 + \frac{1}{2}M_*\dot{\bld{R}}_*^2 \\ &
\begin{split} - \int\! d\bld{x}\ &\bigg\{ \rho\bigg[ \bzeta\cdot\frac{\partial\bld{a}}{\partial t} + \frac{\partial\phi}{\partial t} + \frac{1}{2}\bld{u}\cdot\bld{u} + E\left(\rho,S(\bld{a})\right) + \Psi \\ &- \frac{GM_h}{\left|\bld{x}-\bld{R}\right|} \bigg] + \frac{1}{8\pi G}\nabla\Psi\cdot\nabla\Psi - \frac{\partial\rho}{\partial t}\bld{x}\cdot\dot{\bld{R}}_* \bigg\},
\end{split}
\end{split}
\end{equation}
where $\bld{R}\equiv\bld{R}_h-\bld{R}_*$. In the centre-of-mass frame of the system, (\ref{eq:lag1}) becomes
\begin{equation} \label{eq:lag2}
\begin{split}
L &= \frac{1}{2}\mu\dot{\bld{R}}^2 \\ &
\begin{split} - \int\! &d\bld{x}\ \bigg\{ \rho\bigg[ \bzeta\cdot\frac{\partial\bld{a}}{\partial t} + \frac{\partial\phi}{\partial t} + \frac{1}{2}\bld{u}\cdot\bld{u} + E\left(\rho,S(\bld{a})\right) + \Psi \\ &- \frac{GM_h}{\left|\bld{x}-\bld{R}\right|} \bigg] + \frac{1}{8\pi G}\nabla\Psi\cdot\nabla\Psi + \frac{M_h}{M}\frac{\partial\rho}{\partial t}\bld{x}\cdot\dot{\bld{R}} \bigg\},
\end{split}
\end{split}
\end{equation}
where $M\equiv M_h+M_*$ is the total mass, and $\mu\equiv M_hM_*/M$ is the reduced mass.
\subsection{Homentropic potential flow}
The variation of (\ref{eq:lag2}) with respect to $\bld{a}$ gives us
\begin{equation} \label{eq:genzeta}
\frac{\partial\bzeta}{\partial\tau} = \left(\frac{\partial E}{\partial S}\right)\left(\frac{\partial S}{\partial\bld{a}}\right).
\end{equation}
We shall now derive the conditions on $\bzeta$ for a homentropic potential flow. By definition, the velocity field for a potential flow has the representation $\bld{u} = \nabla \chi'$ for some arbitrary scalar potential $\chi'$. We now define a new potential $\chi$ such that $\chi' = \chi + \phi$. The velocity field can then be written as
\begin{equation*}
\bld{u} = \nabla\chi + \nabla\phi = \frac{\partial\chi}{\partial a_i}\nabla a_i + \nabla\phi.
\end{equation*}
Comparing this expression with (\ref{eq:genvel}), we find that for a potential flow
\begin{equation} \label{eq:potzeta}
\zeta_i = \frac{\partial\chi}{\partial a_i}.
\end{equation}
We now note that for a homentropic fluid it follows from (\ref{eq:genzeta}) that $\chi(\bld{a},\tau) = \chi_1(\bld{a})+\chi_2(\tau)$. Therefore, after substituting (\ref{eq:potzeta}), we find that the Lagrangian (\ref{eq:lag2}) takes the form
\begin{equation*}
\begin{split}
L &= \frac{1}{2}\mu\dot{\bld{R}}^2 \\ &
\begin{split} - \int\! &d\bld{x}\ \bigg\{ \rho\bigg[ \frac{\partial(\chi_1+\phi)}{\partial t} + \frac{1}{2}[\nabla(\chi_1+\phi)]^2 + E(\rho) + \Psi \\ & - \frac{GM_h}{\left|\bld{x}-\bld{R}\right|} \bigg] + \frac{1}{8\pi G}\nabla\Psi\cdot\nabla\Psi + \frac{M_h}{M}\frac{\partial\rho}{\partial t}\bld{x}\cdot\dot{\bld{R}} \bigg\}.
\end{split}
\end{split}
\end{equation*}
Since $\chi_1$ and $\phi$ only appear as the combination $\chi_1+\phi$, we can re-define $\chi_1+\phi\to\phi$ to obtain
\begin{equation} \label{eq:lag}
\begin{split}
L &= \frac{1}{2}\mu\dot{\bld{R}}^2 \\ &
\begin{split} - \int\! &d\bld{x}\ \bigg\{ \rho\bigg[ \frac{\partial\phi}{\partial t} + \frac{1}{2}\nabla\phi\cdot\nabla\phi + E(\rho) + \Psi \\ & - \frac{GM_h}{\left|\bld{x}-\bld{R}\right|} \bigg] + \frac{1}{8\pi G}\nabla\Psi\cdot\nabla\Psi + \frac{M_h}{M}\frac{\partial\rho}{\partial t}\bld{x}\cdot\dot{\bld{R}} \bigg\}.
\end{split}
\end{split}
\end{equation}
It follows from our method of construction that all homentropic potential flows can be derived from (\ref{eq:lag}) with $\phi$ as the velocity potential.

We now note that the conservative nature of gravitational forces guarantees that, in the absence of viscosity, tidal excitation will not generate vorticity. Therefore, if the fluid starts out with zero vorticity then the flow will always be irrotational. We can therefore use (\ref{eq:lag}) for problems involving non-rotating, homentropic stars (e.g.~cold white dwarfs). From here onwards we shall assume this to be the case.
\section{Equations of Motion and Conservation Laws} \label{sec:equations}
\subsection{The equations of motion}
We consider the first-order perturbations to $\phi$, $\Psi$, and $\rho$:
\begin{align*}
\phi(\bld{x},t) &= \phi_0(r) + \phi_1(\bld{x},t), \\
\Psi(\bld{x},t) &= \Psi_0(r) + \Psi_1(\bld{x},t), \\
\rho(\bld{x},t) &= \rho_0(r) + \rho_1(\bld{x},t),
\end{align*}
where $r\equiv|\bld{x}|$, and we have assumed a static, spherically symmetric unperturbed fluid configuration. Retaining terms up to quadratic order in the perturbations, we can separate the Lagrangian as
\begin{equation} \label{eq:lagpieces}
L = L_{\rm eq} + L_{\rm int} + L_{\rm pert},
\end{equation}
where $L_{\rm eq}$ describes the equilibrium (unperturbed) configuration, $L_{\rm pert}$ describes the structure of the perturbations, and $L_{\rm int}$ describes the orbit-perturbation interaction. After making the perturbative expansion, we obtain
\begin{align}
&\begin{aligned}L&_{\rm eq} = \frac{1}{2} \mu \dot{\bld{R}}^2 + \frac{G M_h M_*}{R} \\ & - \int\! d\bld{x}\  \left[ \rho_0 E_0 + \rho_0\Psi_0 + \frac{1}{8\pi G} \left(\frac{d\Psi_0}{dr}\right)^2 \right],\end{aligned} \\
&\begin{aligned}L_{\rm int} = -\int\! d\bld{x}\ \bigg\{ &\rho_0\left(\dot{\phi}_1 + \Psi_1\right) \\ &+ \rho_1\left(h_0 + \Psi_0 - \frac{GM_h}{|\bld{x}-\bld{R}|}\right) \\ &+ \frac{1}{4\pi G}\frac{d\Psi_0}{dr}\frac{\partial\Psi_1}{\partial r}  + \frac{M_h}{M}\dot{\rho}_1\bld{x}\cdot\dot{\bld{R}} \bigg\},\end{aligned} \\
&\begin{aligned}L_{\rm pert} = -&\int\! d\bld{x}\ \bigg\{ \rho_1\left(\dot{\phi}_1 + \Psi_1\right) + \frac{1}{2}\frac{c_s^2}{\rho_0}\rho_1^2 \\ &+ \frac{1}{2}\rho_0\nabla\phi_1\cdot\nabla\phi_1 + \frac{1}{8\pi G}\nabla\Psi_1\cdot\nabla\Psi_1 \bigg\},\end{aligned}
\end{align}
where $R\equiv |\bld{R}|$, and $h_0$ and $c_s$ are the unperturbed specific enthalpy and adiabatic sound speed, respectively. The variations of $L_{\rm eq}$  with respect to $\bld{R}$, $\rho_0$, and $\Psi_0$ yield
\begin{align}
&\mu\ddot{\bld{R}} = - \frac{G M_h M_*}{R^2} \hat{\bld{R}}, \label{eq:kepler} \\
&h_0+\Psi_0 = 0, \\
&\frac{1}{r^2}\frac{d}{dr}\left(r^2\frac{d\Psi_0}{dr}\right) = 4\pi G \rho_0, \label{eq:poisson}
\end{align}
where $\hat{\bld{R}}$ is a unit vector. Together with an equation of state for the fluid, (\ref{eq:kepler})--(\ref{eq:poisson}) determine the unperturbed configuration.

To determine the equations for the perturbations, we consider the variations of $L$ with respect to $\phi_1$, $\rho_1$ and $\Psi_1$. These give us
\begin{align}
&\dot{\rho}_1 + \nabla\cdot\left(\rho_0\nabla\phi_1\right) = 0, \label{eq:linmasscons2} \\
&\dot{\phi}_1 + \frac{c_s^2}{\rho_0}\rho_1 + \Psi_1 = \frac{GM_h}{|\bld{x}-\bld{R}|} + \frac{M_h}{M}\bld{x}\cdot\ddot{\bld{R}}, \label{eq:linbern2} \\
&\nabla^2\Psi_1 = 4\pi G\rho_1. \label{eq:linpoisson2}
\end{align}
It is straightforward to show that, with $M_h=0$, (\ref{eq:linmasscons2})--(\ref{eq:linpoisson2}) are the conventional equations for the normal modes of a non-rotating, homentropic star (see Appendix \ref{app:modes}).

The equation for the orbit, including the back-reaction of the perturbations, is obtained by considering the variation of $L$ with respect to $\bld{R}$:
\begin{equation} \label{eq:orbit}
\mu\ddot{\bld{R}} = -\frac{G M_h M_*}{R^2}\hat{\bld{R}} + G M_h \frac{\partial}{\partial\bld{R}}\int\! d\bld{x}\ \frac{\rho_1}{|\bld{x}-\bld{R}|},
\end{equation}
where we have used the fact that
\begin{equation*}
\int\! d\bld{x}\ \rho_1\bld{x} = 0
\end{equation*}
(this is equivalent to choosing the origin of the coordinates $\bld{x}$ to be the centre-of-mass of the fluid).
\subsection{Amplitude formulation}
It is convenient to write the equations for tidal excitation in terms of the normal mode amplitudes. Let $\bxi$ be the physical displacement field of fluid elements within the star. Integrating (\ref{eq:linmasscons2}) with respect to time and setting the integration constant to zero, we get
\begin{equation} \label{eq:linmasscons3}
\rho_1 = -\nabla\cdot (\rho_0\bxi).
\end{equation}
Taking the gradient of (\ref{eq:linbern2}), we have
\begin{equation}
\ddot{\bxi} + \nabla\left(\frac{c_s^2}{\rho_0}\rho_1 + \Psi_1\right) = GM_h\nabla\left(\frac{1}{|\bld{x}-\bld{R}|}\right) + \frac{M_h}{M}\ddot{\bld{R}}. \label{eq:radialdisp}
\end{equation}
The second term on the left side of (\ref{eq:radialdisp}) can be written as a linear, spatial operator on $\bxi$:
\begin{equation}
\mathcal{D}(\bxi) \equiv \nabla\left[ -\frac{c_s^2}{\rho_0}\nabla\cdot\left(\rho_0\bxi\right) + G\int\! d\bld{x'}\ \frac{\nabla'\cdot(\rho_0'\bxi')}{\left|\bld{x'}-\bld{x}\right|} \right] ,
\end{equation}
where we have used (\ref{eq:linmasscons3}) and (\ref{eq:linpoisson2}) to write $\rho_1$ and $\Psi_1$ in terms of $\bxi$. We can solve for the eigenfunctions $\hat{\bxi}_j(\bld{x})$:
\begin{equation*}
\mathcal{D}(\hat{\bxi}_j) = \omega_{j}^2\hat{\bxi}_j,
\end{equation*}
where $\omega_j^2$ are the eigenvalues. It can be proved that the operator $\mathcal{D}$ is Hermitian with respect to mass \citep{cha64,cox80}. We are therefore guaranteed that its eigenvalues are real, and it is generally assumed that its eigenfunctions form a complete set and can be made mutually orthogonal. Hence, with the appropriate choice of normalization,
\begin{equation} \label{eq:xiortho}
\int\! d\bld{x}\ \rho_0\hat{\bxi}_{j'}^*\cdot\hat{\bxi}_j = \delta_{jj'}.
\end{equation}
Expanding $\bxi$ as
\begin{equation} \label{eq:xiexpand}
\bxi (\bld{x},t) = \sum_j A_j(t)\hat{\bxi}_j(\bld{x})
\end{equation}
in (\ref{eq:radialdisp}) and then projecting out a single mode gives
\begin{equation} \label{eq:sho}
\ddot{A}_j + \omega_j^2A_j = f_j(\bld{R}) + \frac{M_h}{M}\ddot{\bld{R}}\cdot\int\! d\bld{x}\ \rho_0\hat{\bxi}_j^*,
\end{equation}
where
\begin{equation} \label{eq:overlap}
f_j(\bld{R}) \equiv GM_h\int\! d\bld{x}\ \rho_0\hat{\bxi}_j^*\cdot\nabla\left(\frac{1}{|\bld{x}-\bld{R}|}\right)
\end{equation}
(see Appendix~\ref{app:overlap}). This is just a forced, harmonic oscillator with frequency $\omega_j$. The second term on the right side of (\ref{eq:sho}) is non-zero only for for monopolar ($l=0$) modes, and exactly cancels the first term for that case. This is a mathematical statement of the fact that monopolar modes are not tidally excited.

The equation for the orbit (\ref{eq:orbit}), after substituting (\ref{eq:linmasscons3}) and then performing an integration by parts, becomes
\begin{align*}
\mu\ddot{\bld{R}} &= -\frac{G M_h M_*}{R^2}\hat{\bld{R}} \\ &+ G M_h \frac{\partial}{\partial\bld{R}}\int\! d\bld{x}\ \rho_0\bxi\cdot\nabla\left(\frac{1}{|\bld{x}-\bld{R}|}\right).
\end{align*}
Expanding $\bxi$ in terms of the normal modes, we find
\begin{equation}
\mu\ddot{\bld{R}} = -\frac{G M_h M_*}{R^2}\hat{\bld{R}} + \sum_j A_j^*\frac{\partial}{\partial\bld{R}} f_j(\bld{R}).
\end{equation}
\subsection{Energy and angular momentum}
To find the conserved energy $\mathcal{E}$, we calculate the time-time component of the energy-momentum tensor from (\ref{eq:lagpieces}) using
\begin{equation} \label{eq:enmomtensor}
T^{i}_{\ j} = \frac{\partial\mathcal{L}}{\partial(\partial_{i}q_k)}\partial_{j}q_k - \delta^{i}_{\ j}\mathcal{L},
\end{equation}
where $\mathcal{L}$ is the Lagrangian density, and $q_k$ are the generalized fields. A straightforward evaluation gives
\begin{equation}  \label{eq:energy}
\begin{split}
\mathcal{E} &= \int\! d\bld{x}\ T^t_{\ t} \\ &= \frac{1}{2}\mu\dot{\bld{R}}^2 - \frac{G M_h M_*}{R} - G M_h\int\! d\bld{x}\ \frac{\rho_1}{|\bld{x}-\bld{R}|} \\  &+ \frac{1}{2}\int\! d\bld{x}\ \Bigg( \rho_0\nabla\phi_1\cdot\nabla\phi_1 + \frac{c_s^2}{\rho_0}\rho_1^2 + \rho_1\Psi_1 \Bigg),
\end{split}
\end{equation}
where we have used (\ref{eq:linpoisson2}) to eliminate the gravitational self-energy of the perturbations. The total energy is the sum of three components: orbital, perturbation, and coupling. The various pieces are easily identified in (\ref{eq:energy}).

We can, without loss of generality, assume the orbit to be in the equatorial plane. The conserved angular momentum $L_z$ can then be calculated from the appropriate component of the energy-momentum tensor as
\begin{equation} \label{eq:angmom}
\begin{split}
L_z &= \int\! d\bld{x}\ T^t_{\ \varphi} \\ &= \mu R^2\dot{u} - \int\! d\bld{x}\ \rho_1\frac{\partial\phi_1}{\partial\varphi},
\end{split}
\end{equation}
where $\varphi$ and $u$ are the azimuthal coordinates associated with $\bld{x}$ and $\bld{R}$, respectively.

It may be noted that the canonical form (\ref{eq:enmomtensor}) of the energy-momentum tensor is not manifestly symmetric. However, it is well-known that the tensor can be made symmetric by the addition of a suitable divergence term:
\begin{equation*}
{T'}^{ij} = T^{ij} + \frac{\partial}{\partial x^k}\psi^{ijk}, \qquad \psi^{ijk} = -\psi^{ikj}.
\end{equation*}
We do not need to do this since integral quantities such as (\ref{eq:energy}) and (\ref{eq:angmom}) are unaffected by such a transformation \citep{lan75}.

The energy and angular momentum associated with the normal modes also take on relatively simple forms in terms of the time-dependent amplitudes $A_j$. From (\ref{eq:energy}), we know that the energy associated with perturbations is
\begin{equation} \label{eq:pertenergy}
\mathcal{E}_{\rm pert} = \frac{1}{2}\int\! d\bld{x}\ \left[ \rho_0\dot{\bxi}\cdot\dot{\bxi} + \rho_1\left(\frac{c_s^2}{\rho_0}\rho_1 + \Psi_1\right) \right].
\end{equation}
Substituting (\ref{eq:linmasscons3}) into (\ref{eq:pertenergy}) and then integrating the second term in the integrand by parts gives
\begin{equation*}
\mathcal{E}_{\rm pert} = \frac{1}{2}\int\! d\bld{x}\ \rho_0\left[ \dot{\bxi}\cdot\dot{\bxi} + \bxi\cdot\nabla\left(\frac{c_s^2}{\rho_0}\rho_1 + \Psi_1\right) \right].
\end{equation*}
Note that the second term in the integrand now involves the same linear operator that we used to define the normal modes. Expanding $\bxi$ as in (\ref{eq:xiexpand}) and using the orthonormality relation (\ref{eq:xiortho}), we find that the energy associated with mode $j$ is just
\begin{equation} \label{eq:energy2}
\mathcal{E}_j = \frac{1}{2}\left(|\dot{A}_j|^2 + \omega_j^2|A_j|^2\right).
\end{equation}
From (\ref{eq:angmom}) and (\ref{eq:linmasscons3}), we know that the angular momentum associated with perturbations is
\begin{equation*}
L_z^{\rm pert} = \int\! d\bld{x}\ \nabla\cdot\left(\rho_0\bxi\right)\frac{\partial\phi_1}{\partial\varphi}.
\end{equation*}
Performing an integration by parts, we get
\begin{equation}
L_z^{\rm pert} = -\int\! d\bld{x}\ \rho_0\bxi\cdot\frac{\partial\dot{\bxi}}{\partial\varphi}.
\end{equation}
Once again expanding $\bxi$ as in (\ref{eq:xiexpand}), using the orthonormality relation (\ref{eq:xiortho}), and the fact that
\begin{equation*}
\frac{\partial\hat{\bxi}_j}{\partial\varphi} = im\hat{\bxi}_j
\end{equation*}
(see Appendix \ref{app:modes}), we find that the angular momentum associated with mode $j$ is just
\begin{equation} \label{eq:angmom2}
L_j = -i m \dot{A}_j A_j^*.
\end{equation}
We shall now derive a simple relation between the energy and angular momentum associated with an isolated mode (i.e.~with $f_j=0$). In that case, from (\ref{eq:sho}) we have $A_j(t)\propto e^{i\omega_jt}$. Therefore, from (\ref{eq:energy2}) and (\ref{eq:angmom2}) we get
\begin{equation} \label{eq:rogerstheorem}
\frac{\mathcal{E}_j}{L_j} = \frac{\omega_j^2|A_j|^2}{m\omega_j|A_j|^2} = \frac{\omega_j}{m}.
\end{equation}
This relation is to be expected on physical grounds as follows. If we consider tidally exciting the mode at resonance in a circular orbit, then the rate at which energy is transferred to the mode is just
\begin{equation*}
\frac{d\mathcal{E}_j}{dt} = \tau\Omega = \frac{dL_j}{dt}\frac{\omega_j}{m},
\end{equation*}
where $\tau$ is the torque exerted by the perturbing mass, and $\Omega$ is the orbital frequency. Integrating this equation with respect to time and setting the initial mode energy and angular momentum to zero, we obtain (\ref{eq:rogerstheorem}).

Finally, the conserved energy (\ref{eq:energy}) and angular momentum (\ref{eq:angmom}), written in terms of the amplitudes, are
\begin{equation}
\begin{aligned}
\mathcal{E} = \frac{1}{2}\mu\dot{\bld{R}}^2 &- \frac{G M_h M_*}{R} - \sum_j A_j^*f_j(\bld{R}) \\ &+ \frac{1}{2}\sum_j\left(|\dot{A}_j|^2 + \omega_j^2|A_j|^2\right),
\end{aligned}
\end{equation}
and
\begin{equation}
L_z = \mu R^2\dot{u} - \sum_j i m \dot{A}_j A_j^*.
\end{equation}
\subsection{Post-Newtonian corrections} \label{sec:postnewt}
For problems involving compact binaries, general relativistic effects may be important. The corrections to the orbit are partly periodic, but there are important secular effects, such as periastron advance and gravitational radiation reaction, which must be taken into account for even a qualitatively accurate treatment of the problem. The most straightforward scheme, and the one we adopt, is to treat the corrections as perturbations to the Newtonian orbit, neglecting all finite-size effects. In this scheme, we simply add the appropriate corrections to the orbital acceleration, energy and angular momentum at each time-step. Detailed derivations of these corrections may be found elsewhere \citep[cf.][]{iye95}. We only list the relevant equations here for reference as they will be used in a forthcoming publication. The equations are taken from \citet{iye95}.

The first- and second-order corrections to the orbital accelerations are
\begin{align}
&\begin{aligned}\ddot{\bld{R}}_{\rm PN}^{(1)} &= -\frac{GM}{R^2}\bigg\{\bigg[-2(2+\eta)\frac{GM}{c^2R} + (1+3\eta)\frac{\dot{\bld{R}}^2}{c^2} \\ &- \frac{3}{2}\eta\frac{\dot{R}^2}{c^2}\bigg]\hat{\bld{R}} - 2(2-\eta)\frac{\dot{R}\dot{\bld{R}}}{c^2}\bigg\}, \end{aligned} \\
&\begin{aligned}\ddot{\bld{R}}_{\rm PN}^{(2)} &= -\frac{GM}{R^2}\bigg\{\bigg[\frac{3}{4}(12+29\eta)\left(\frac{GM}{c^2r}\right)^2 \\ &+ \eta(3-4\eta)\frac{\dot{\bld{R}}^4}{c^4} + \frac{15}{8}\eta(1-3\eta)\frac{\dot{R}^4}{c^4} \\ &- \frac{3}{2}\eta(3-4\eta)\frac{\dot{R}^2\dot{\bld{R}}^2}{c^4} - \frac{1}{2}\eta(13-4\eta)\frac{GM\dot{\bld{R}}^2}{c^4R} \\ &-(2+25\eta+2\eta^2)\frac{GM\dot{R}^2}{c^4R}\bigg]\hat{\bld{R}} \\ &- \frac{1}{2}\bigg[\eta(15+4\eta)\frac{\dot{\bld{R}}^2}{c^2} - (4+41\eta+8\eta^2)\frac{GM}{c^2R} \\ &- 3\eta(3+2\eta)\frac{\dot{R}^2}{c^2}\bigg]\frac{\dot{R}\dot{\bld{R}}}{c^2}\bigg\}, \end{aligned}
\end{align}
where $\eta\equiv\mu/M$. These corrections are formally conservative in the sense that it is still possible to define a conserved energy and a conserved angular momentum to second-order in $\dot{\bld{R}}^2/c^2$. The appropriate corrections to the Newtonian quantities are
\begin{align}
&\begin{aligned}\mathcal{E}_{\rm PN}^{(1)} &= \mu c^2\bigg[\frac{3}{8}(1-3\eta)\frac{\dot{\bld{R}}^4}{c^4}+\frac{1}{2}(3+\eta)\frac{GM\dot{\bld{R}}^2}{c^4R} \\ &+\frac{1}{2}\eta\frac{GM\dot{R}^2}{c^4R} + \frac{1}{2}\left(\frac{GM}{c^2R}\right)^2\bigg], \end{aligned} \\
&\begin{aligned}\mathcal{E}_{\rm PN}^{(2)} &= \mu c^2\bigg[\frac{5}{16}(1-7\eta+13\eta^2)\frac{\dot{\bld{R}}^6}{c^6} \\ &+ \frac{1}{8}(21-23\eta-27\eta^2)\frac{GM\dot{\bld{R}}^4}{c^6R} \\ &+ \frac{1}{4}\eta(1-15\eta)\frac{GM\dot{R}^2\dot{\bld{R}}^2}{c^6R} - \frac{3}{8}\eta(1-3\eta)\frac{GM\dot{R}^4}{c^6R} \\ &- \frac{1}{4}(2+15\eta)\left(\frac{GM}{c^2R}\right)^3 \\ &+ \frac{1}{8}(14-55\eta+4\eta^2)\left(\frac{GM\dot{\bld{R}}}{c^3R}\right)^2 \\ &+ \frac{1}{8}(4+69\eta+12\eta^2)\left(\frac{GM\dot{R}}{c^3R}\right)^2\bigg], \end{aligned} \\
\end{align}
and
\begin{align}
&\begin{aligned}L_{\rm PN}^{(1)} = \mu R^2\dot{u}\bigg[\frac{1}{2}(1-3\eta)\frac{\dot{\bld{R}}^2}{c^2} + (3+\eta)\frac{GM}{c^2R}\bigg], \end{aligned} \\
&\begin{aligned}L_{\rm PN}^{(2)} &= \mu R^2\dot{u}\bigg[\frac{1}{2}(7-10\eta-9\eta^2)\frac{GM\dot{\bld{R}}^2}{c^4R} \\ &- \frac{1}{2}\eta(2+5\eta)\frac{GM\dot{R}^2}{c^4R} \\ &+ \frac{1}{4}(14-41\eta+4\eta^2)\left(\frac{GM}{c^2R}\right)^2 \\ &+ \frac{3}{8}(1-7\eta+13\eta^2)\frac{\dot{\bld{R}}^4}{c^4}\bigg]. \end{aligned}
\end{align}
At $5/2$ post-Newtonian order, the first dissipative terms arise. Therefore, after second-order it is no longer possible to define a conserved energy or angular momentum. The $5/2$- and $7/2$-order corrections to the orbital acceleration are given by
\begin{equation}
\begin{aligned}
\ddot{\bld{R}}_{\rm RR} = -\frac{8}{5}\eta\left(\frac{GM}{R^2}\right)\left(\frac{GM}{c^2R}\right)\bigg[&-(A_{5/2}+A_{7/2})\frac{\dot{R}}{c}\hat{\bld{R}} \\ &+ (B_{5/2}+B_{7/2})\frac{\dot{\bld{R}}}{c}\bigg],
\end{aligned}
\end{equation}
where
\begin{align}
&\begin{aligned}A_{5/2} = 18\frac{\dot{\bld{R}}^2}{c^2} + \frac{2}{3}\frac{GM}{c^2R} - 25\frac{\dot{R}^2}{c^2}, \end{aligned} \\
&\begin{aligned}B_{5/2} = 6\frac{\dot{\bld{R}}^2}{c^2} - 2\frac{GM}{c^2R} - 15\frac{\dot{R}^2}{c^2}, \end{aligned} \\
&\begin{aligned}A_{7/2} &= \left(\frac{87}{14}-48\eta\right)\frac{\dot{\bld{R}}^4}{c^4} - \left(\frac{5379}{28}+\frac{136}{3}\eta\right)\frac{GM\dot{\bld{R}}^2}{c^4R} \\ &+ \frac{25}{2}(1+5\eta)\frac{\dot{R}^2\dot{\bld{R}}^2}{c^4} + \left(\frac{1353}{4}+133\eta\right)\frac{GM\dot{R}^2}{c^4R} \\ &- \frac{35}{2}(1-\eta)\frac{\dot{R}^4}{c^4} + \left(\frac{160}{7}+\frac{55}{3}\eta\right)\left(\frac{GM}{c^2R}\right)^2, \end{aligned} \\
&\begin{aligned}B_{7/2} &= -\frac{27}{14}\frac{\dot{\bld{R}}^4}{c^4} - \left(\frac{4861}{84}+\frac{58}{3}\eta\right)\frac{GM\dot{\bld{R}}^2}{c^4R} \\ &+ \frac{3}{2}(13-37\eta)\frac{\dot{R}^2\dot{\bld{R}}^2}{c^4} + \left(\frac{2591}{12}+97\eta\right)\frac{GM\dot{R}^2}{c^4R} \\ &- \frac{25}{2}(1-7\eta)\frac{\dot{R}^4}{c^4} + \frac{1}{3}\left(\frac{776}{7}+55\eta\right)\left(\frac{GM}{c^2R}\right)^2. \end{aligned}
\end{align}

\section{Applicability}
Our formalism, as developed in the previous section, is valid only for non-rotating, homentropic stars. However, the Lagrangian (\ref{eq:lag2}) has more general applicability to any problem where dissipative effects may be neglected. It is therefore possible, in principle, to generalise the formalism to include rotating fluid configurations and non-homentropic equations of state. In this section we shall discuss further limitations of the formalism developed in \S~\ref{sec:equations}.

The two most important cases where our formalism is not applicable are regimes of strong non-linear effects and gravity. In particular, one fundamental assumption upon which our mode analysis is based is that the underlying fluid configuration remains spherically symmetric. This is no longer the case, for example, when the star fills its Roche lobe and is tidally disrupted. We can formulate this as a quantitative limit using an approximation to the effective Roche lobe radius due to \citet{egg83}:
\begin{equation}
\frac{R_*}{R} < \frac{0.49\,q^{2/3}}{0.6\,q^{2/3} + \ln\left(1+q^{1/3}\right)},
\end{equation}
where $q\equiv M_*/M_h$.

The limitation in the regime of strong gravitational fields is due to the post-Newtonian corrections summarised in \S~\ref{sec:postnewt}. In particular, the perturbation series in the order parameter $\dot{\bld{R}}^2/c^2$, represented by the corrections, fails to provide a satisfactory approximation in the strong-field limit. The point at which the post-Newtonian corrections fail is roughly at $R=10\,GM/c^2$ \citep{bra98}. Higher order corrections have been calculated \citep{gop97}, however they are not expected to significantly increase the region of validity of our formalism. To evolve the orbit in the strong-field regime, a fully relativistic treatment of the two-body problem is required.

\section{Conclusion}
In this paper, we have described and developed a variational approach for handling a specific dynamical problem---the oscillation-orbit interaction of a non-rotating star with a compact object companion. Our treatment is presented in some generality because we believe that this method can also be employed in other problems associated with accretion disks and extra-solar planets. We recover the standard, Newtonian equations of fluid dynamics for irrotational flow as wells as normal mode equations for small oscillations. The power of the Lagrangian approach is made manifest in the identification of algebraic expressions for the modal energy and angular momentum, and for the equations of motion. It also suggests an approach to handling more complex problems, including those involving strong-field gravity and rotational flow.

In a forthcoming paper, we will use this approach to compute the dynamical evolution of a white dwarf in orbit about a compact object as it spirals inward under the action of gravitational radiation.

\section*{Acknowledgments}
We would like to thank Alessandra Buonanno, Donald Lynden-Bell, and Kaloyan Penev for several useful discussions. We also acknowledge support under NASA grant NAGWS-2837.
\bibliography{lagrangian_formalism}

\appendix
\section{The Euler Equation} \label{app:euler}
We shall assume, for simplicity, that the centre-of-mass frame of the fluid is inertial. Then, in the centre-of-mass frame, the Lagrangian (\ref{eq:flulag1}) becomes
\begin{equation*}
L_* = \int\! d\bld{a}\ \Big[ \frac{1}{2}\left(\frac{\partial\bld{x}}{\partial\tau}\right)^2 - E\left(\alpha,S(\bld{a})\right) - \Phi(\bld{x}) \Big],
\end{equation*}
where $\alpha\equiv\rho^{-1}$ is the specific volume. The variation with respect to $\bld{x}$ yields
\begin{equation} \label{eq:euler_inter1}
\begin{split}
\frac{\partial^2x_i}{\partial\tau^2} &- \frac{\partial E}{\partial\alpha}\frac{\partial}{\partial a_j}\left[\frac{\partial\alpha}{\partial(\partial x_i/\partial a_j)}\right] \\ &- \frac{\partial\alpha}{\partial(\partial x_i/\partial a_j)}\frac{\partial}{\partial a_j}\left(\frac{\partial E}{\partial\alpha}\right) = -\frac{\partial\Phi}{\partial x_i}.
\end{split}
\end{equation}
We can show that the second term on the left hand side of (\ref{eq:euler_inter1}) is zero as follows. From the definition of $\alpha$, we know that
\begin{equation*}
\alpha = \frac{\partial(\bld{x})}{\partial(\bld{a})} = \frac{1}{6}\epsilon_{ijk}\epsilon_{lmn}\frac{\partial x_i}{\partial a_l}\frac{\partial x_j}{\partial a_m}\frac{\partial x_k}{\partial a_n} ,
\end{equation*}
where $\epsilon_{ijk}$ is the three-dimensional Levi-Civita symbol \citep[cf.][]{arf95}.
Taking the derivative of both sides with respect to $\partial x_i/\partial a_j$, we get
\begin{equation*}
\frac{\partial\alpha}{\partial(\partial x_i/\partial a_j)} = \frac{1}{2}\epsilon_{ijk}\epsilon_{lmn}\frac{\partial x_j}{\partial a_m}\frac{\partial x_k}{\partial a_n} .
\end{equation*}
And, finally, taking a derivative with respect to $a_j$ gives us
\begin{equation*}
\frac{\partial}{\partial a_j}\left[\frac{\partial\alpha}{\partial(\partial x_i/\partial a_j)}\right] = \epsilon_{ijk}\epsilon_{lmn}\frac{\partial^2x_j}{\partial a_l\partial a_m}\frac{\partial x_k}{\partial a_n} = 0.
\end{equation*}
Next, using the definition of $\alpha$ and the identity (\ref{eq:matrixinv}), we note that
\begin{equation*}
\frac{\partial\alpha}{\partial(\partial x_i/\partial a_j)}\frac{\partial}{\partial a_j} = \alpha\frac{\partial a_j}{\partial x_i}\frac{\partial}{\partial a_j} = \alpha\frac{\partial}{\partial x_i}.
\end{equation*}
Therefore, (\ref{eq:euler_inter1}) now becomes
\begin{equation*}
\frac{\partial^2x_i}{\partial\tau^2} = -\alpha\frac{\partial P}{\partial x_i} - \frac{\partial\Phi}{\partial x_i},
\end{equation*}
where $P\equiv-\partial E/\partial\alpha$ is the pressure. Recalling that $\partial/\partial\tau$ corresponds to a convective derivative in the Eulerian description, we see that this is just the Euler equation.

\section{The Normal Modes} \label{app:modes}
We expand $\phi_1$, $\rho_1$, $\Psi_1$ and $|\bld{x}-\bld{R}|^{-1}$ in terms of spherical harmonics:
\begin{align*}
\phi_1(\bld{x},t) &= \sum_{l,m}\phi_{lm}(r,t)Y_{lm}(\hat{\bld{x}}), \\
\rho_1(\bld{x},t) &= \sum_{l,m}\rho_{lm}(r,t)Y_{lm}(\hat{\bld{x}}), \\
\Psi_1(\bld{x},t) &= \sum_{l,m}\Psi_{lm}(r,t)Y_{lm}(\hat{\bld{x}}), \\
\frac{1}{|\bld{x}-\bld{R}|} &= \sum_{l,m}\frac{4\pi}{2l+1}\frac{r^l}{R^{l+1}}Y_{lm}^*(\hat{\bld{R}})Y_{lm}(\hat{\bld{x}})
\end{align*}
\citep[cf.][]{jac99}. After inserting the expansions and integrating over angles, (\ref{eq:linmasscons2})--(\ref{eq:linpoisson2}) become
\begin{align}
&\dot{\rho}_{lm} + \frac{1}{r^2}\frac{\partial}{\partial r}\left(r^2\rho_0\frac{\partial\phi_{lm}}{\partial r}\right) - \frac{l(l+1)}{r^2}\rho_0\phi_{lm} = 0, \label{eq:linmasscons} \\
&\begin{aligned}\dot{\phi}_{lm} + \frac{c_s^2}{\rho_0}\rho_{lm} + \Psi_{lm} = &-\frac{G M_h}{R}\frac{4\pi}{2l+1} \left(\frac{r}{R}\right)^lY_{lm}^*(\hat{\bld{R}}) \\ & + \delta_{l1}\frac{4\pi}{3}\frac{M_h}{M}r|\ddot{\bld{R}}| Y_{lm}^*(\hat{\ddot{\bld{R}}}), \end{aligned} \label{eq:linbern} \\
&\frac{1}{r^2}\frac{\partial}{\partial r}\left(r^2\frac{\partial\Psi_{lm}}{\partial r}\right) - \frac{l(l+1)}{r^2}\Psi_{lm} = 4\pi G\rho_{lm}. \label{eq:linpoisson}
\end{align}
The first term on the right hand side of (\ref{eq:linbern}) is a forcing term that couples the modes to the gravitational potential of the point-mass. The second term, which is only present for dipolar ($l=1$) modes, exactly cancels the first term for that case. We can therefore conclude that dipolar modes are not tidally excited. This was to be expected since the origin of the coordinates $\bld{x}$ is the centre-of-mass of the fluid.

The temporal Fourier transforms of (\ref{eq:linmasscons})--(\ref{eq:linpoisson}), with $M_h=0$, give
\begin{align}
&\left[\frac{1}{r^2}\frac{d}{dr}r^2\rho_0\frac{d}{dr} - \frac{l(l+1)}{r^2}\rho_0\right]\tilde{\phi}_{lm} = -i\omega\tilde{\rho}_{lm}, \label{eq:fourlinmasscons} \\
&i\omega\tilde{\phi}_{lm} + \frac{c_s^2}{\rho_0}\tilde{\rho}_{lm} + \tilde{\Psi}_{lm} = 0, \label{eq:fourlinbern} \\
&\left[\frac{1}{r^2}\frac{d}{dr}r^2\frac{d}{dr} - \frac{l(l+1)}{r^2}\right]\tilde{\Psi}_{lm} = 4\pi G\tilde{\rho}_{lm}, \label{eq:fourlinpoisson}
\end{align}
where $\tilde{\rho}_{lm}$, $\tilde{\phi}_{lm}$ and $\tilde{\Psi}_{lm}$ are the temporal Fourier transforms of $\rho_{lm}$, $\phi_{lm}$, and $\Psi_{lm}$, respectively. We can use (\ref{eq:fourlinbern}) to eliminate $\tilde{\rho}_{lm}$ from (\ref{eq:fourlinmasscons}) and (\ref{eq:fourlinpoisson}). This yields the two second-order equations
\begin{align}
&\left[\frac{1}{r^2\rho_0}\frac{d}{dr}r^2\rho_0\frac{d}{dr}-\frac{l(l+1)}{r^2}+\frac{\omega^2}{c_s^2}\right]\tilde{\phi}_{lm} = i\frac{\omega}{c_s^2}\tilde{\Psi}_{lm}, \label{eq:operator1} \\
&\left[\frac{1}{r^2}\frac{d}{dr}r^2\frac{d}{dr}-\frac{l(l+1)}{r^2}+\frac{4\pi G\rho_0}{c_s^2}\right]\tilde{\Psi}_{lm} = -i\frac{4\pi G\rho_0\omega}{c_s^2}\tilde{\phi}_{lm}. \label{eq:operator2}
\end{align}
With the definitions
\begin{align*}
&\tilde{\eta}_1 \equiv \frac{1}{i\omega r}\frac{d\tilde{\phi}_{lm}}{dr}, &&\tilde{\eta}_2 \equiv \frac{\omega}{igr}\tilde{\phi}_{lm}, \\
&\tilde{\eta}_3 \equiv \frac{1}{gr}\tilde{\Psi}_{lm}, &&\tilde{\eta}_4 \equiv \frac{1}{g}\frac{d\tilde{\Psi}_{lm}}{dr},
\end{align*}
and
\begin{align*}
&U \equiv \frac{d\ln\mathcal{M}}{d\ln r}, &&V \equiv -\frac{d\ln P_0}{d\ln r}, \\
&\Gamma_1 \equiv \frac{d\ln P_0}{d\ln\rho_0}, &&C \equiv \frac{M_*}{\mathcal{M}}\left(\frac{r}{r_*}\right)^3, \\
&\sigma^2 \equiv \frac{r_*^3}{G M_*}\omega^2,
\end{align*}
where
\begin{align*}
&\mathcal{M}(r) \equiv \int_0^{r}\! dr'\ 4\pi {r'}^2\rho_0(r'), \\
&g(r) \equiv \frac{G\mathcal{M}(r)}{r^2},
\end{align*}
and after some manipulation, (\ref{eq:operator1}) and (\ref{eq:operator2}) give the four first-order equations
\begin{align}
&r\frac{d\tilde{\eta}_1}{dr} = \left(\frac{V}{\Gamma_1}-3\right)\tilde{\eta}_1 + \left[\frac{l(l+1)}{\sigma^2C}-\frac{V}{\Gamma_1}\right]\tilde{\eta}_2 + \frac{V}{\Gamma_1}\tilde{\eta}_3, \label{eq:eta1} \\
&r\frac{d\tilde{\eta}_2}{dr} = \sigma^2C\tilde{\eta}_1 + (1-U)\tilde{\eta}_2, \label{eq:eta2} \\
&r\frac{d\tilde{\eta}_3}{dr} = (1-U)\tilde{\eta}_3 + \tilde{\eta}_4, \label{eq:eta3} \\
&r\frac{d\tilde{\eta}_4}{dr} = \frac{UV}{\Gamma_1}\tilde{\eta}_2 + \left[l(l+1)-\frac{UV}{\Gamma_1}\right]\tilde{\eta}_3 - U\tilde{\eta}_4. \label{eq:eta4}
\end{align}
These are the conventional equations for the normal mode structure of a non-rotating, homentropic star \citep[cf.][]{cox80,kip90}.

To have a well-posed problem, we obviously need to specify four boundary conditions. At the center of the star, we require that the variables $\tilde{\eta}_i$ be well-behaved. Expanding in a power series around $r=0$, we have
\begin{equation*}
\tilde{\eta}_i = \sum_{\alpha=0}^{\infty}A^{(i)}_{\alpha}r^{\alpha}.
\end{equation*}
Substituting into (\ref{eq:eta1})--(\ref{eq:eta4}), and using the fact that
\begin{align*}
&\lim_{r\rightarrow 0} U = 3, \\
&\lim_{r\rightarrow 0} V = 0, \\
&\lim_{r\rightarrow 0} C = \rm{constant},
\end{align*}
we find that the only non-vanishing coefficients correspond to $\alpha = l-2$, and
\begin{align}
l\tilde{\eta}_2 &= \sigma^2C\tilde{\eta}_1, \label{eq:bc1} \\
\tilde{\eta}_4 &= l\tilde{\eta}_3, \label{eq:bc2}
\end{align}
which constitute our boundary conditions at the center. At the surface, we require that $\Psi_{lm}$ satisfy the Laplace equation. This gives
\begin{equation}
\tilde{\eta}_4 = -(l+1)\tilde{\eta}_3, \label{eq:bc3}
\end{equation}
at $r=r_*$, as our third boundary condition. Finally, from (\ref{eq:eta1}) and using the condition that
\begin{equation*}
\lim_{r\rightarrow r_*} V = \frac{GM_*}{r_*} \lim_{r\rightarrow r_*} \frac{\rho_0}{P_0} = \infty
\end{equation*}
\citep[cf.][]{cox80}, we get a fourth boundary condition that
\begin{equation}
\tilde{\eta}_2 = \tilde{\eta}_1 + \tilde{\eta}_3, \label{eq:bc4}
\end{equation}
at the surface. Together with the boundary conditions (\ref{eq:bc1})--(\ref{eq:bc4}), (\ref{eq:eta1})--(\ref{eq:eta4}) consitute an eigenvalue problem for the normal modes.

We now turn to the problem of determining the physical displacement of fluid elements from the unperturbed configuration in terms of the $\tilde{\eta}_i$. Expanding $\tilde{\phi}_1$ (the temporal Fourier transform of the perturbation to $\phi$) in terms of the normal modes, we have
\begin{equation*}
\tilde{\phi}_1 = \sum_{n,l,m}\tilde{A}_{nlm}(\omega)\tilde{\phi}_{nlm}(r)Y_{lm}(\hat{\bld{x}}),
\end{equation*}
where $\tilde{\phi}_{nlm}$ is a normalized eigenfunction, and we use the subscript $n$ to distinguish between the various modes corresponding to the same $l,m$. Using the fact that $\dot{\bxi}=\nabla\phi_1$, where $\bxi$ is the physical displacement of fluid elements, we find
\begin{equation} \label{eq:xifourtran}
\tilde{\bxi}(\bld{x},\omega) = \sum_{n,l,m} \tilde{A}_{nlm}(\omega)\hat{\bxi}_{nlm}(\bld{x}),
\end{equation}
where
\begin{equation*}
\hat{\bxi}_{nlm}(\bld{x}) \equiv \left(r\tilde{\eta}_1\hat{\bld{x}}+ \frac{r^2}{\sigma^2C}\tilde{\eta}_2\nabla\right) Y_{lm}(\hat{\bld{x}}).
\end{equation*}
The displacement field $\bxi(\bld{x},t)$ is just the temporal inverse Fourier transform of (\ref{eq:xifourtran}).

Note that by taking the gradient of (\ref{eq:linbern2}), it is straightforward to see that the normal modes $\hat{\bxi}_{nlm}$ satisfy the operator equation
\begin{equation*}
\left[\nabla\left(\frac{c_s^2}{\rho_0}\rho_1 + \Psi_1\right)\right](\hat{\bxi}_{nlm}) = \omega_{nlm}^2\hat{\bxi}_{nlm}.
\end{equation*}

\section{The Overlap Integral} \label{app:overlap}
Integrating (\ref{eq:overlap}) by parts and substituting (\ref{eq:linmasscons3}), we find
\begin{equation*}
f_j(\bld{R}) = G M_h\int\! d\bld{x}\ \frac{\rho_j^*(\bld{x})}{|\bld{x}-\bld{R}|}.
\end{equation*}
From (\ref{eq:linpoisson2}), we know that
\begin{equation*}
\Psi_j^*(\bld{R}) = -G \int\! d\bld{x}\ \frac{\rho_j^*(\bld{x})}{|\bld{x}-\bld{R}|}.
\end{equation*}
Therefore, we have
\begin{equation} \label{eq:overlap2}
f_j(\bld{R}) = -M_h\Psi_j^*(\bld{R}).
\end{equation}
It is straightforward to write $\Psi_j(\bld{R})$ in terms of the dimensionless variable $\tilde{\eta}_3$ defined in Appendix~\ref{app:modes} using the solution to the Laplace equation in spherical coordinates:
\begin{equation*}
\Psi_j(\bld{R}) = \frac{GM_*}{R_*}\tilde{\eta}_3(R_*)\left(\frac{R_*}{R}\right)^{l+1} Y_{lm}(\hat{\bld{R}}).
\end{equation*}
Therefore, the overlap integral (\ref{eq:overlap2}) is
\begin{equation*}
f_j(\bld{R}) = -\frac{GM_hM_*}{R_*}\tilde{\eta}_3(R_*)\left(\frac{R_*}{R}\right)^{l+1} Y_{lm}^*(\hat{\bld{R}}).
\end{equation*}
The above expression for the overlap integral explicitly demonstrates that in the absence of a perturbation to the gravitational field there is no mode excitation. This is, of course, a direct consequence of Newton's third law. Assuming the orbit to be in the equatorial plane, the expression becomes
\begin{equation}
\begin{aligned}
f_j(\bld{R}) = -\frac{GM_hM_*}{R_*}&\tilde{\eta}_3(R_*)Y_{lm}\left(\frac{\pi}{2},0\right) \\ &\times\left(\frac{R_*}{R}\right)^{l+1}e^{i m u},
\end{aligned}
\end{equation}
where $u$ is the azimuthal coordinate associated with $\bld{R}$.

\bsp
\label{lastpage}
\end{document}